\documentclass[prl, twocolumn, showpacs, showkeys]{revtex4}%
\usepackage{amsfonts}
\usepackage{amsmath}
\usepackage{amssymb}
\usepackage{psfrag}
\usepackage{graphicx}%
\begin{document}
\preprint{ }
\title[ ]{Transconductance of a double quantum dot system in the Kondo regime}
\author{V. Koerting}
\email[author to whom correspondence should be addressed:]
      {koerting@tkm.physik.uni-karlsruhe.de}
\author{P. W\"{o}lfle}
\affiliation{Institut f\"{u}r Theorie der Kondensierten Materie, Universit\"{a}t Karlsruhe,
D-76128 Karlsruhe, Germany}
\author{J. Paaske}
\affiliation{The Niels Bohr Institute \& Nano-Science Center, University of Copenhagen,
DK-2100 Copenhagen, Denmark}


\begin{abstract}
We consider a lateral double-dot system in the Coulomb blockade
regime with a single spin-1/2 on each dot, mutually coupled by an
anti-ferromagnetic exchange interaction.
Each of the two dots is contacted by two leads. We demonstrate
that the voltage across one of the dots will have a profound
influence on the current passing through the other dot. Using Poor
Man's scaling, we find that the Kondo-effect can lead to a
strong enhancement of this {\it transconductance}.
\end{abstract}

\date{\today}
%\received{} \revised{} \accepted{}

\maketitle

%%%%%%%%%%%%%%%%%%%%%%%%%%%%%%%%%%%%%%%%%%%%%%%%%%%%%%%%%%%%%%%%%%%
%%%%%%%%%%%%%%%%%%%% Introduction %%%%%%%%%%%%%%%%%%%%%%%%%%%%%%%%%
%%%%%%%%%%%%%%%%%%%%%%%%%%%%%%%%%%%%%%%%%%%%%%%%%%%%%%%%%%%%%%%%%%%

Nano-structured electron systems offer the unique possibility to
prepare and probe highly nonequilibrium stationary states. Energy
dissipation and therefore heating, characteristic of any
nonequilibrium state, takes place in the bulk reservoirs connected
to the nano-structure, where it is absorbed causing a negligible
change of state. The nature of such a nonequilibrium state may be
conveniently probed by transport measurements, rather than
thermodynamic or optical experiments (cf.~e.g.~Refs.\cite{Aleiner:02,Wiel:03}).
In an ordinary quantum dot set-up the
probing field, e.g.~the applied bias-voltage, is also responsible
for driving the system out of equilibrium. However, by separating
the driving field and the probing field, many more possibilities for
both studying nonequilibrium states and exploiting their properties
arise. In the simplest realization of such a system one must have
two parts of the system strongly coupled by some interaction and
each contacted by two leads. Applying a sufficiently high bias
across one part will then influence the current through the other
part. This coupling is mediated by the change in nonequilibrium
occupation numbers on the nano-structure caused by the driving
field, which in turn governs the current-response to the probing
field.

\begin{figure}[t]
\centering
\includegraphics[width = 0.35 \textwidth]{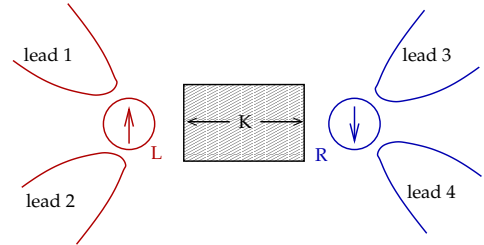}
\caption{(color online) Double Quantum
Dot System: Two Kondo impurities denoted $L$, $R$ each connected to
two leads $1, 2$ and $3, 4$, respectively, and coupled to each
other by a spin exchange interaction $K$.}%
\label{fig:device_new}%
\end{figure}

In this letter we investigate the properties of a system of this
type (cf.~Fig.~\ref{fig:device_new}), that is two quantum dots in the
Coulomb blockade regime with odd electron occupation, giving rise to
spins $S_{L,R}=1/2$, coupled by an exchange interaction $K$. Each
dot is contacted by two leads, modeled by exchange tunneling
amplitudes $J_{L,R}\sim t^2/E_C$, in terms of lead-dot tunneling
amplitudes $t$ and charging energy $E_C$.
This setup is very similar to the double-dot devices studied
for example in Refs.~\cite{Craig:04, Sasaki:06, Jeong:01}, where the
central region was either a quantum-dot, a quantum-wire or a
tunneling barrier, which could then support an exchange interaction
between the two spins.
The following discussion does not rely on the physical nature of the
exchange coupling. In practice, it can be due to a simple
superexchange mechanism~\cite{Golovach:03} or an RKKY-interaction,
as is most likely the case in Refs.~\cite{Craig:04, Sasaki:06}, or
even due to the so-called Kondo-doublet interaction suggested
recently~\cite{Simonin:06}.

The theory of the two-impurity Kondo model (2IKM) per se is well
established~\cite{2IKM}, but with the experimental progress in
quantum-dot systems, issues like the quantum critical properties
remain an active field of research~\cite{Zarand:06,Chung:06,Varma:06}.
The influence of the exchange-interaction $K$ and an applied
magnetic field on the nonlinear conductance $dI_L/dV_L$ was studied
in Refs.~\cite{Vavilov:05,Simon:05}, and it was demonstrated how the
nature of the ground-state ({\it Kondo-screened} or {\it local
singlet}) can be extracted from the temperature dependence of the
linear conductance.

In this present setup with two sets of biased leads, we are facing
an entirely new problem in which correlations and nonequilibrium
effects must be treated side by side.
We calculate the transconductance $dI_L/dV_R$ and demonstrate a
transistor-effect in which a current through the left dot, $I_L$, as
a response to a bias on the left dot, $V_L$, is switched on only
when the bias over the right dot, $V_R$, exceeds a threshold given
by $K>0$. This threshold effect rests on the fact that the
exchange-tunneling current from lead 1 to lead 2, $I_L$, is zero as
long as the spins are locked in a singlet. A sufficiently large bias
across the right dot will, however, lead to a partial occupation of
the triplet state, which does allow for a finite current $I_L$. As
we shall demonstrate, the nonequilibrium polarization due to $V_R$,
and thereby the transconductance peak at $V_R\sim K$, will be
enhanced by the Kondo-effect.
In the case of ferromagnetic coupling, the triplet ground state
gives rise to a spin-$1$ Kondo-effect~\cite{Vavilov:05, Simon:05}.
In this case, a bias-voltage over the right quantum dot will
influence the corresponding zero-bias conductance peak and singlet
satellite peaks in $dI_L/dV_L$,
partly by decoherence and partly by a redistribution of the
singlet/triplet occupations. However, these effects are less
dramatic and we therefore restrict our attention to the
antiferromagnetic case, $K>0$, which exhibits the afore mentioned
threshold effect.

%%%%%%%%%%%%%%%%%%%%%%%%%%%%%%%%%%%%%%%%%%%%%%%%%%%%%%%%%%%%%%%%%%%
%%%%%%%%%%%%%%%%%%%% Model %%%%%%%%%%%%%%%%%%%%%%%%%%%%%%%%%%%%%%%%
%%%%%%%%%%%%%%%%%%%%%%%%%%%%%%%%%%%%%%%%%%%%%%%%%%%%%%%%%%%%%%%%%%%

{\it The model.} We model the double dot by two spin $\frac{1}{2}$
operators $\mathbf{S}_{L},\mathbf{S}_{R}$ mutually coupled by
exchange interaction $K$ and each coupled to a pair of leads
(couplings $J_{L,R}>0$) as sketched in Fig.~\ref{fig:device_new}.
The Hamiltonian reads
\begin{align}
H=&\sum_{{\bf k}n\sigma}(\epsilon_{k}-\mu_{n})
c^{\dagger}_{n{\bf k}\sigma}c_{n{\bf k}\sigma}
  +K\ \mathbf{S}_{L}\cdot\mathbf{S}_{R}\\
 &+{\textstyle\sum\limits_{n,m=1,2}}J_{L}^{nm}\ \mathbf{S}_{L}
\cdot\mathbf{s}_{nm}+{\textstyle\sum\limits_{n,m=3,4}}J_{R}^{nm}\
\mathbf{S}_{R}\cdot\mathbf{s}_{nm},\nonumber
\end{align}
where $n,m=1,2,3,4$ labels the leads, assumed to be in equilibrium
at chemical potentials $\mu_{n}$.
$c_{n{\bf k}\sigma}^{\dagger}$ creates electrons in lead $n$ of
momentum ${\bf k}$ and spin $\sigma$ and the spin-operator
$\mathbf{s}_{nm}
= \sum_{{\bf k},{\bf k}^{\prime},\sigma,\sigma^{\prime}} c_{n{\bf k}\sigma}^{\dagger}
(\vec{\tau}_{\sigma,\sigma\prime}/2) c_{m{\bf k}^{\prime}\sigma^{\prime}}$
denotes respectively the conduction
electron spin ($n=m$) and the exchange tunneling operator ($n\neq
m$). We do not allow for charge transfer between the two dots.
Experimentally the charge transfer may be suppressed by a suitable
arrangement of tunneling barriers~(cf.~\cite{Zarand:06}).

For $K=0$, the two spins will undergo separate screening by the
conduction electron spins in each their set of leads as the
temperature is lowered beneath the two, generally different,
Kondo-temperatures $T_{K,L}$ and $T_{K,R}$. However, at sufficiently
large bias voltages $eV_{L} \gg T_{K,L}$ and/or $eV_{R} \gg T_{K,R}$ where
$eV_{L}, eV_{R}$ are the chemical potential differences of leads
$n=1,2$ or leads $n=3,4$, respectively, these Kondo effects are
partially suppressed \cite{Kaminski:00,Rosch:01}. A finite
anti-ferromagnetic exchange coupling, $K$, on the other hand, will
correlate the spins $\mathbf{S}_{L},\mathbf{S}_{R}$, and depending
on the relative magnitude of $K$ and the Kondo-temperatures, the two
spins will either lock into a singlet or be screened by their
respective lead electrons~\cite{2IKM}. We will be interested in the
regime $K \gg \max\{T_{K,L},T_{K,R}\}$, when the Kondo effect cannot
develop fully. In this case it is necessary to treat the exchange
coupling $K$ exactly, whereas it is possible to treat the
Kondo effect perturbatively.

The spin-states of the isolated double-dot are the singlet ($S=0$)
and triplet ($S=1$) states of the two coupled spin 1/2. They form
the appropriate basis states in this regime even when the coupling
to the leads is turned on. For convenience, we employ a pseudo-boson
({\it bond-operator}~\cite{Sachdev:90}) representation of the four
impurity-states, with creation operators
$\{b_{\gamma}^{\dagger}\}=(s^{\dagger},t_{+}^{\dagger},t_{0}^{\dagger},
t_{-}^{\dagger})$, i.e. $\gamma\in\{s,+,0,-\}$.
The energy eigenvalues of the four states are
$\omega_{s}=-\frac{3}{4}K$, $\omega_{0}=\omega_{+,-}=\frac{1}{4}K=\omega_{t}$,
and therefore $K\ \mathbf{S}_{L}\cdot\mathbf{S}_{R}
=\sum_{\gamma}\ \omega_{\gamma}b_{\gamma}^{\dagger}b_{\gamma}$, subjected to the
constraint $Q=s^{\dagger}s+t_{0}^{\dagger}t_{0}+t_{+}^{\dagger}t_{+}
+t_{-}^{\dagger}t_{-}=1$. The constraint may be exactly imposed
by adding a term $\lambda Q$ to the Hamiltonian and taking the limit
$\lambda \rightarrow\infty$~\cite{Abrikosov:65}. In terms of the
pseudo-bosons, the spin-1/2 operators are given by
$S_{L,R}^{z}=\frac{1}{2}(\pm s^{\dagger}t_{0}\pm t_{0}^{\dagger}s
+t_{+}^{\dagger}t_{+}-t_{-}^{\dagger}t_{-})$ and
$S_{L,R}^{+}=\frac{1}{\sqrt{2}}\left(  S_{L,R}^{x}+iS_{L,R}^{y}\right)
=(S_{L,R}^{-})^{\dagger}=\frac{1}{2}(\pm s^{\dagger}t_{-}\mp t_{+}^{\dagger}s
+t_{+}^{\dagger}t_{0}+t_{0}^{\dagger}t_{-})$~\cite{Sachdev:90}, or in compact
notation $\mathbf{S}_{L,R}=\frac{1}{2}\sum_{\gamma,\gamma^{\prime}}\
b_{\gamma}^{\dagger}\mathbf{T}_{L,R;\gamma\gamma^{\prime}}b_{\gamma^{\prime}}$,
where $\mathbf{T}_{L,R;\gamma\gamma^{\prime}}$ is a vector of three
4x4 matrices, $T^{x},T^{y},T^{z}$. Notice that this
exchange-tunneling model contains no operators of the form
$s^{\dagger}s$, which immediately implies that the current $I_L$ is
zero unless transitions to the triplet-states become energetically
allowed.

It is known that away from the particle-hole symmetric point a
finite potential scattering term will give rise to a current below
the threshold. Nevertheless, we shall assume that a set of
individual gates on each dot can be adjusted so as to make the
system (nearly) particle-hole symmetric and hence neglect potential
scattering altogether. In the setup by Sasaki {\it et
al.}~\cite{Sasaki:06}, such tuning was made possible by two
individual plunger-gates.

\textit{Nonequilibrium perturbation theory.} In equilibrium, i.e.
for vanishing bias voltages, the states $|\gamma\rangle$ of the
double dot are thermally occupied, with occupation numbers
$n_{\gamma}=\langle b_{\gamma}^{\dagger}b_{\gamma}\rangle
=\exp(-\omega_{\gamma}/T)/\sum_{\gamma^{\prime}}\exp(-\omega_{\gamma^{\prime}}/T)$.
In a stationary nonequilibrium situation with sufficiently large currents
through the nano-structure, the occupation numbers are determined by
the currents, i.e. the couplings and the voltages, rather than
temperature~\cite{pertKondo}.
A further complication is that nondiagonal parts of the density matrix
$n_{\gamma\gamma^{\prime}}=\langle b_{\gamma}^{\dagger}b_{\gamma^{\prime}}\rangle$
may be nonzero due to the mixing of states by the coupling to the
leads. Confining our considerations to zero magnetic field this will
not be the case and furthermore the triplet-states will be equally
occupied, i.e. $n_{\gamma}=n_{t}$, for $\gamma=+,0,-$. It is useful
to define the polarization $p = n_s - n_t$, and we note that
$\langle {\bf S}_L\cdot{\bf S}_R \rangle = -\frac{3}{4} p$.

The nonequilibrium occupation-numbers are determined from the
steady-state quantum Boltzmann equation,
$\sum_{\gamma^{\prime}\neq\gamma}\left[\Gamma_{\gamma\gamma^{\prime}}n_{\gamma^\prime}
-\Gamma_{\gamma^{\prime}\gamma}n_{\gamma}\right]=0$,
with transition-rates given by the golden rule expression
\begin{equation}
\Gamma_{\gamma\gamma^{\prime}}\!=\!\frac{\pi}{4}
{\textstyle\sum\limits_{n,m}} \mathbf{T}_{\gamma^\prime\gamma}^{mn}
\cdot\mathbf{T}_{\gamma\gamma^\prime}^{nm}
\!{\textstyle\int\limits_{-\infty}^{\infty}}\!\!d\omega
g_{\gamma\gamma^\prime}^{nm}(\omega)
g_{\gamma^{\prime}\gamma}^{mn}(\omega-\omega_{\gamma^{\prime}\gamma})
F_{\gamma\gamma^\prime}^{nm}(\omega),\label{eq:def_gamma}
\end{equation}
where $(n,m)$ takes the values $(1,2)$ or $(3,4)$ and
$\mathbf{T}^{nm} = \mathbf{T}_L$ or $\mathbf{T}_R$,
$\omega_{\gamma\gamma^\prime}=\omega_{\gamma}-\omega_{\gamma^\prime}$
and $F_{\gamma\gamma^\prime}^{nm}(\omega)=f(\omega-\mu_{n})
[1-f(\omega-\mu_{m}-\omega_{\gamma\gamma^\prime})]$, $f(\omega)$ being the
Fermi-function. The dimensionless exchange-couplings,
$g^{nm}=\nu_{F}J^{nm}$, (assuming same density of states, $\nu_{F}$,
in all leads) will only acquire frequency-dependence under
renormalization as we shall discuss later. The Boltzmann equation is
readily solved together with the constraint
$\sum_{\gamma}n_{\gamma}=1$, and using the three-fold degeneracy of
the triplet, one finds that $p = (\Gamma_{st} -
\Gamma_{ts})/(\Gamma_{st} + 3 \Gamma_{ts})$.

%%%%%%%%%%%%%%%%%%%%%%%%%%%%%%%%%%%%%%%%%%%%%%%%%%%%%%%%%%%%%%%%%%%%%
%%%%%%%%%%%%%%%%%%% Current %%%%%%%%%%%%%%%%%%%%%%%%%%%%%%%%%%%%%%%%%
%%%%%%%%%%%%%%%%%%%%%%%%%%%%%%%%%%%%%%%%%%%%%%%%%%%%%%%%%%%%%%%%%%%%%

The current through the left dot is obtained as
\begin{align}
I_{L}(V)=&\frac{\pi^2}{2}\frac{e}{h}
{\textstyle\sum\limits_{\gamma\gamma^{\prime}}}
n_{\gamma}\mathbf{T}_{L;\gamma\gamma^{\prime}}
\cdot\mathbf{T}_{L;\gamma^{\prime}\gamma}\label{eq:defcurrent}\\
&\!\!\!\times\!\!{\textstyle\int\limits_{-\infty}^{\infty}} d\omega
g_{\gamma\gamma^{\prime}}^{21}(\omega)
g_{\gamma^{\prime}\gamma}^{12}(\omega-\omega_{\gamma^{\prime}\gamma})
[F_{\gamma^\prime\gamma}^{12}(\omega)-F_{\gamma^\prime\gamma}^{21}(\omega)],
\nonumber
\end{align}
and a corresponding expression holds for $I_{R}(V)$, with leads
$1,2$ replaced by $3,4$.

%%%%%%%%%%%%%%%%%%%%%%%%%%%%%%%%%%%%%%%%%%%%%%%%%%%%%%%%%%%%%%%%%%%%%
%%%%%%%%%%%%%%%%%%% Transconductance %%%%%%%%%%%%%%%%%%%%%%%%%%%%%%%%
%%%%%%%%%%%%%%%%%%%%%%%%%%%%%%%%%%%%%%%%%%%%%%%%%%%%%%%%%%%%%%%%%%%%%

From Eq.~\eqref{eq:defcurrent} it is clear that $V_R$ only enters
via the occupation numbers, and using the unrenormalized couplings
the transconductance is given by
\begin{align}
\frac{dI_L}{dV_R} &= \frac{6\pi^2 e}{h K} g_c^2
p^2\left[F_{-}(V_L)-eV_L\right]\frac{d}{dV_R}F_{+}(V_R),
\label{eq:transcond}
\end{align}
where $F_{\pm}(V)=(K-eV)n_{B}(K-eV)\pm(K+eV)n_{B}(K+eV)$, in terms of
the Bose-function $n_B(\varepsilon)=1/(e^{\varepsilon/T}-1)$, and
$g_c^2=(g^{12}g^{34})^2/\sum_{m, n}(g^{mn})^2$. $dI_L/dV_R$ is an odd
function of $V_R$ which diverges like $T/(K-eV_R)$ for $eV_R\to K$. As
we shall see below, this divergence is removed once the broadening
of the excited triplet-states is taken into account, and one is left
with a pronounced peak in the transconductance. Assuming that
$g^{mn}=g_{L(R)}$ for ${m,n}\in\{1,2\}, \{3,4\}$, one has
$g_c^2=(g_L^{-2}+g_R^{-2})^{-1}$ which shows that the
transconductance peak is expected to be largest in a symmetric setup
where $g_L=g_R$.

%%%%%%%%%%%%%%%%%%%%%%%%%%%%%%%%%%%%%%%%%%%%%%%%%%%%%%%%%%%%%%%%%%%%%
%%%%%%%%%%%%%%%%%%% RG %%%%%%%%%%%%%%%%%%%%%%%%%%%%%%%%%%%%%%%%%%%%%%
%%%%%%%%%%%%%%%%%%%%%%%%%%%%%%%%%%%%%%%%%%%%%%%%%%%%%%%%%%%%%%%%%%%%%

\textit{Renormalized perturbation theory.} Since the higher order
terms in the perturbation series are plagued by logarithmic
singularities characteristic of the Kondo effect, we apply Poor
man's scaling to sum up the leading logarithms in this series. As
demonstrated earlier~\cite{Rosch:01, pertRG}, this is
conveniently done by keeping track of the generated frequency
dependence of the couplings. Within this approach the renormalized
coupling {\it functions} are calculated from the following set of
renormalization group (RG) equations describing the flow of couplings
under a reduction of the bandwidth D:
\begin{align}
\frac{\partial g_{tt}^{nm}(\omega)}{\partial\ln D}  &  =-\frac{1}%
{2}{\textstyle\sum\limits_{l}}\{g_{st}^{lm}(\mu_{l})g_{ts}^{nl}(\mu
_{l}-K)\Theta_{\omega-\mu_{l}+K}\nonumber\\
&  \qquad\left.  +g_{st}^{nl}(\mu_{l}+K)g_{ts}^{lm}(\mu_{l})\Theta_{\omega
-\mu_{l}-K}\right. \nonumber\\
&  \qquad+2g_{tt}^{lm}(\mu_{l})g_{tt}^{nl}(\mu_{l})\Theta_{\omega-\mu_{l}}\} \\
\frac{\partial g_{ts}^{nm}(\omega)}{\partial\ln D}  &  =-{\textstyle\sum
\limits_{l}}\{g_{ts}^{lm}(\mu_{l})g_{tt}^{nl}(\mu_{l})\Theta_{\omega-\mu_{l}%
}\nonumber\\
&  \qquad+g_{ts}^{nl}(\mu_{l}-K)g_{tt}^{lm}(\mu_{l})\Theta_{\omega-\mu_{l}%
+K}\},
\end{align}
with $g_{st}^{nm}(\omega)=g_{ts}^{mn}(\omega - K)$. Here
$\Theta_{\omega}=\Theta(D^{2}-\omega^{2}-\Gamma^{2})$ is the step
function and $\Gamma$ is a Korringa-like decoherence-rate due to
particle-hole excitations in the leads. In general, $\Gamma$
involves both self-energy, and vertex corrections and may depend on
the intermediate state for which it describes the broadening
(cf.~\cite{Paaske_deph}). Since, however, $\Gamma$ appears only
under the logarithm we will neglect these details here and simply
use the maximum $\Gamma=\mathrm{max}\{ \Gamma_{\gamma\gamma^\prime}
\}$ with $\Gamma_{\gamma\gamma^{\prime}}$ defined in
Eq.~\eqref{eq:def_gamma}. The renormalized couplings and $\Gamma$
are calculated self-consistently, then we solve for the occupation
numbers and calculate the current using the renormalized
coupling-functions in Eq.~\eqref{eq:defcurrent}. The perturbative RG
approach is valid for $K \gg T_K$, where $T_K = D_0 e^{-1/2 g(D_0)}$
in terms of bare coupling $g(D_0)$ and the half-bandwidth $D_0$.

%%%%%%%%%%%%%%%%%%%%%%%%%%%%%%%%%%%%%%%%%%%%%%%%%%%%%%%%%%%%%%%%%%%%%
%%%%%%%%%%%%%%%%%%% Plots and discussion %%%%%%%%%%%%%%%%%%%%%%%%%%%%
%%%%%%%%%%%%%%%%%%%%%%%%%%%%%%%%%%%%%%%%%%%%%%%%%%%%%%%%%%%%%%%%%%%%%

\begin{figure}[t]
\vspace{0.75 cm}
\centering
\includegraphics[width = 0.35 \textwidth]{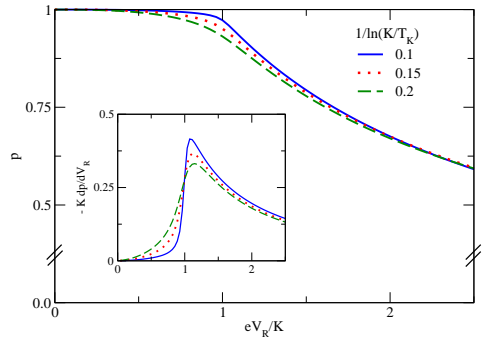}
\caption{(color online) Polarization $p = n_s - n_t$ versus voltage $eV_R/K$ for
different values of $K/T_K$ at $T/T_K = 0.1$ and voltage $V_L = 0$.
Inset shows the derivative of the curves in the main panel.}
\label{fig:pol}%
\end{figure}

\textit{Renormalized conductances.} In Fig.~\ref{fig:pol} we plot
the polarization $p$ as a function of $eV_{R}/K$ at $V_{L}=0$, for
several values of $K/T_{K}$ and at temperature $T/T_{K}=0.1$.
As $eV_{R}$ approaches $K$ the triplet states become partially
populated and eventually $p$ tends to zero as $1/V_R$. Since we use
$T\ll K$, temperature is practically zero and the broadening is
governed by $K$ and $V_R$. In the inset of Fig.~\ref{fig:pol} we
plot the derivative $-K dp/dV_R$ vs. $eV_R/K$, showing a pronounced
peak which is increasingly smeared as the ratio $K/T_K$ is made
smaller and the Kondo-correlations become more pronounced.

\begin{figure}[ht]
\vspace{0.75 cm} \centering
\includegraphics[width = 0.35 \textwidth]{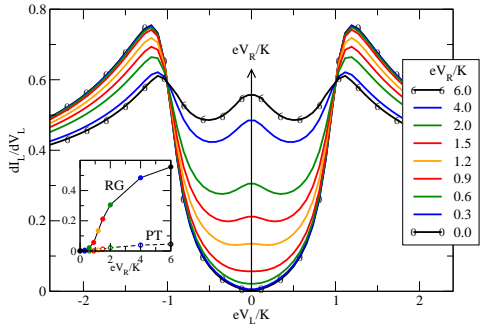}
\caption{(color online)
Differential conductance $dI_L/dV_L$ in units of $e^2/h$
for fixed $T/T_K = 0.1$ and $1/\ln(K/T_K)=0.2$
versus $eV_L/K$ at different values of
$eV_R/K$. Inset: $dI_L/dV_L$ at $V_L \rightarrow 0$ versus $eV_R/K$,
full circles: perturbative RG, empty circles: $2^\mathrm{nd}$ order
PT.}
\label{fig:GL}%
\end{figure}

From the polarization $p$ and the renormalized couplings, we can now
determine the current from Eq.~\eqref{eq:defcurrent}. As mentioned
above, electron transport via exchange-tunneling always involves
excitation out of the singlet ground state. This requires an energy
of the order of $K$ and therefore leads to a threshold behavior near
$eV_{L} \ll K$, as sketched also in Ref.~\cite{Vavilov:05}.
Alternatively, the singlet-triplet gap can be overcome by a finite
$V_R$, which will lead to a finite $dI_L/dV_L$ even for $eV_L<K$.
This is clearly seen in Fig.~\ref{fig:GL}.
When $V_R$ becomes larger than $K$, a zero-bias peak
is seen to develop, the magnitude of which is determined by the ratio
$\Gamma K/T_{K}^{2}$.
In the limit of zero probing-voltage, we find that
\begin{equation}
\lim_{V_L\to 0}\frac{dI_L}{dV_L}=\frac{6 \pi^2 e^2}{h}
     \frac{n_t}{[ 4 \ln( \sqrt{\Gamma K} / T_K) ]^2},
     \label{eq:dILdVL}
\end{equation}
where both $n_t$ and $\Gamma$ depend on the driving-voltage $V_R$.
The inset of Fig.~\ref{fig:GL} shows the increase of the linear
conductance with $V_{R}$, as calculated respectively by perturbative
RG and $2^\mathrm{nd}$ order perturbation theory~(PT). The initial
rise with $V_R$ is set by the triplet-occupation number
$n_t=(1-p)/4$. The amplitude is logarithmically enhanced due to the
Kondo effect and falls off very slowly ($\Gamma\propto V_R$) for
$V_R\gg K$.

\begin{figure}[t]
\vspace{0.75 cm} \centering
\includegraphics[width = 0.35 \textwidth]{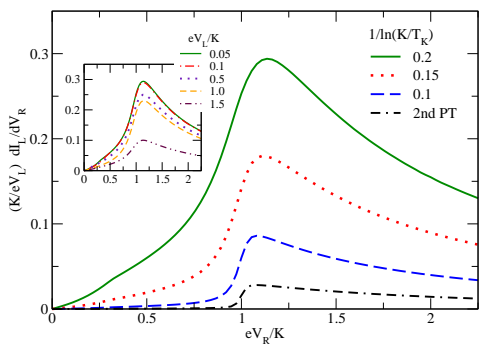}
\caption{(color online) Transconductance $dI_L/dV_R$ in units of
$e^2/h$ multiplied by $(K/eV_L)$ versus $eV_R/K$ for fixed $T/T_K =
0.1$, probing voltage $eV_L/K = 0.05$ and different values of
$K/T_K$. Inset: Transconductance for $1/\ln(K/T_K) = 0.2$ and
different $eV_L/K$.}
\label{fig:trans}%
\end{figure}

Fig.~\ref{fig:trans} shows the transconductance versus $eV_R/K$ for
$V_L \to 0$. Enhancing the Kondo-correlations by lowering $K/T_K$ is
seen to amplify the transconductance but also to smear the
threshold. By scaling with $K/eV_L$ we eliminate the $1/K$
dependence seen in Eq.~\eqref{eq:transcond} as well as most of the
dependence on the probing field for $eV_L < K$~(cf. inset of
Fig.~\ref{fig:trans}).

\textit{Conclusion.} We studied the mutual influence of spatially
separated currents flowing through two quantum-dots coupled only by
spin-exchange. The strong interrelation between the currents and
nonequilibrium occupation numbers for the singlet and triplet
spin-states correlates the two currents, and we showed that this can
be observed as a marked peak in the transconductance. Since the spin
exchange interaction requires well-defined spins on the quantum dots
the coupling to the leads should be relatively small, but opening up
towards the Kondo-regime our perturbative RG calculations 
demonstrate that weak Kondo-correlations ($T_K\ll K$) will in fact
enhance the transconductance peak. This pronounced signal in the
transconductance provides a new experimental means of probing the
exchange-coupling between two quantum-dot spins.
Moreover, as evident from Eq.~(\ref{eq:dILdVL}), this mode of
operation provides direct experimental evidence for the voltage
dependence of the nonequilibrium distribution function $n_t$.

\textit{Acknowledgement}
We acknowledge discussions with Chung-Hou Chung, Lars Fritz and
Matthias Vojta. This work has been supported by the DFG-Center for
Functional Nanostructures (CFN) at the university of Karlsruhe under
project B2.9 (V.K. and P.W.), the Institute for Nanotechnology,
Research Center Karlsruhe (P.W.) and the Danish Agency for Science,
Technology and Innovation (J.P.).

\end{document}